\documentclass[11pt, letter]{amsart}
\usepackage{amsthm,amsmath,amsxtra,amscd,amssymb,xypic,color,xspace,esint}
\usepackage{latexsym,amsmath,amssymb,mathrsfs}
\usepackage{amsfonts,eucal,amsthm,graphicx,color}

\numberwithin{equation}{section}

\newcommand{\ve}{{\varepsilon}}

\newcommand{\cN}{{\mathcal N}}

\def\beq{\begin{equation}}
\def\eeq{\end{equation}}

\theoremstyle{plain}

\theoremstyle{definition}

\newenvironment{aside}{\begin{quote}\sffamily}{\end{quote}}

\newcommand {\BC}   {\mathbb C}

\newcommand {\BR}   {\mathbb R}

\newcommand {\bT}   {\mathbf{T}}

\newcommand {\bQ}   {\mathbf{Q}}
\newcommand {\bR}   {\mathbf{R}}

\newcommand {\qe} {\mathfrak q}

\newcommand {\BZ}   {\mathbb Z}

\newcommand {\CalC} {\mathcal C}

\newcommand {\CalN} {\mathcal N}

\newcommand {\CalP} {\mathcal P}

\newcommand {\CalR} {\mathcal R}
\newcommand {\fR} {\mathfrak{R}}

\newcommand {\cX} {\mathcal{X}}
\newcommand {\cY} {\mathcal{Y}}

\newcommand {\cZ} {\mathcal{Z}}

\newcommand{\fo}{\vert\kern -.03in\_}

\newcommand {\ii} {\mathrm{i}}

\newcommand{\lam}{{\lambda}}

\begin{document}

\title[elliptic CM, instantons, and bilinear relations]{Elliptic Calogero-Moser system, \\
crossed and folded instantons,\\ and bilinear identities}
\author{Andrei Grekov, Nikita Nekrasov}
\address{Simons Center for Geometry and Physics$^{n}$\\
Yang Institute for Theoretical Physics$^{g,n}$\\ 
Stony Brook University, Stony Brook NY 11794-3636, USA}
\date{June 2023}
\begin{abstract}
Affine analogues of the $Q, {\tilde Q}$-functions are constructed using {\em folded instantons} partition functions. They are shown to be the solutions of the quantum spectral curve of the $N$-body elliptic Calogero-Moser (eCM) system, the quantum Krichever curve. $Q, {\tilde Q}$ also solve the elliptic analogue of the quantum Wronskian equation.  In the companion paper we present the quantum analogue of Krichever's Lax operator for eCM. A connection to crossed instantons on Taub-Nut spaces, and opers on a punctured torus is pointed out.
\end{abstract}

\maketitle

\section{Elliptic Calogero-Moser system}

Let 
\beq
{\tau} = {\tau}_{1} + {\ii}{\tau}_{2} \in {\BC}\, ,  \ {\tau}_{2}> 0 \, , \ \qe = e^{2\pi\ii \tau} \ .
\eeq 
Let 
\beq
E_{\tau} = {\BC}/{\BZ}+ {\tau}{\BZ} \approx {\BC}^{\times}/{\qe}^{\BZ}
\eeq
be the corresponding elliptic curve. 

Fix an integer $N \geq 1$, and $n, \hbar \in {\BC}$. Quantum elliptic Calogero-Moser system  (eCM) can be defined in many ways. At simplified level it is a collection of commuting differential operators ${\hat H}_{1}, {\hat H}_{2}, \ldots , {\hat H}_{N}$ in $N$ variables
$z_{1}, \ldots, z_{N}$, 
\beq
\begin{aligned}
& {\hat H}_{1} =  \sum_{i=1}^{N} \frac{\partial}{\partial z_{i}}\, , \\
& {\hat H}_{2} =   - \frac{1}{2} \sum_{i=1}^{N} \frac{\partial^2}{\partial z_{i}^2}\, + \, n ( n - 1) \sum_{i < j} {\wp} ( z_{i} - z_{j} | {\tau} )  \\
& \ldots \\
\end{aligned}
\label{eq:ecmhams}
\eeq
invariant under the action of the double affine Weyl group
\beq
\left( z_{i} \right)_{i=1}^{N} \mapsto \left( z_{{\sigma}(i)} + a_{i} + {\tau} b_{i} \right)_{i=1}^{N}\, , \ {\sigma} \in S(N) \, , \ a_k, b_k \in {\BZ}
\eeq
The elliptic Calogero-Moser system admits several interesting degenerations. One such limit, ${\tau} \to {\ii\infty}$, produces the trigonometric/hyperbolic CM system. Another limit is more subtle: let $z_{k} = y_{k} + {\tau} k/N$, $k  = 1, \ldots , N$, ${\Lambda}^{2N} = {\qe} \left( - ({\hbar} n)^{2} \right)^{N}$. Send $\tau \to \ii\infty, n \to \infty$ while keeping ${\bf y} = (y_{k})_{k=1}^{N}$ and $\Lambda, \hbar$ finite. The result is the $N$-body periodic Toda chain \cite{GuToda}:
\beq
\begin{aligned}
& {\hat H}_{1} =  \sum_{i=1}^{N} \frac{\partial}{\partial y_{i}}\, , \\
& {\hat H}_{2} =   - \frac{1}{2} \sum_{i=1}^{N} \frac{\partial^2}{\partial y_{i}^2}\, + \, ({\Lambda}/{\hbar})^{2} \sum_{i =1}^{N} e^{y_{i} -y_{i+1}} \, ,  \\
& \ldots \\
& y_{N+1} \equiv y_{1} \\
\end{aligned}
\label{eq:todahams}
\eeq

\subsection{Spectral curve of eCM} 
In the classical limit $\hbar \to 0$, $n \to \infty$, ${\hbar n} = {\nu}$ fixed,  the operators \eqref{eq:ecmhams} go over to the Poisson-commuting functions $H_{1}(p,z), \ldots, H_{N}(p,z)$ on ${\CalP}_{\BC} = \left( T^{*}E_{\tau} \right)^{N}/S(N)$, or on its smooth resolution
\beq
{\tilde\CalP}_{\BC} = Hilb^{[N]}(S)\, , 
\label{eq:hilbte}\eeq
the Hilbert scheme of $N$ points on the complex symplectic surface $S$, which is an affine bundle over the cotangent bundle to $E_{\tau}$ (which can be analytically identified with the algebraic torus ${\BC}^{\times} \times {\BC}^{\times}$, or, non-canonically, mapped to $T^{*}E_{\tau}$, isomorphically outside a fiber over one point in $E_{\tau}$). In \cite{Krichever80} the functions $H_{k}$ are constructed with the help of the Lax operator
\beq
{\bf L}(u) = \Vert L_{ij}(u) \Vert_{i,j=1}^{N}\, , \ L_{ij}(u) = p_{i} {\delta}_{ij} + {\nu} (1 - {\delta}_{ij}) \frac{{\theta}(z_{i}- z_{j} + u  | {\tau}) {\theta}^{\prime}(0 | {\tau})}{{\theta}(z_{i}- z_{j} | {\tau}) {\theta}(u | {\tau})}
\eeq
where
\beq
{\theta}(u | {\tau} ) = \sum_{l \in {\BZ}} (-1)^{l} e^{2\pi \ii l u} {\qe}^{\frac{l(l-1)}{2}}\  
\label{eq:thetaf}
\eeq 
obeys
\beq
{\theta}(u+1 | {\tau} ) = {\theta}(u | {\tau} ) = -  e^{2\pi\ii u} {\theta}(u+{\tau} | {\tau} )
\label{eq:thettr}
\eeq 
One defines the {\it spectral curve} ${\CalC} \subset T^{*}\left( E_{\tau} \backslash  \{ u = 0 \} \right)$ via:
\beq
R({\lambda}, u) \equiv {\rm Det} \left( {\lambda} \cdot {\rm Id}_{N} - {\bf L}(u) \right) = 0 
\label{eq:speccurve}
\eeq
As a function of $\lambda$, $R$ is a degree $N$ polynomial, with coefficients meromorphic double-periodic functions of $u$, with poles only at $u =0$. 

An innocent looking change of variables (used in \cite{Krichever80}) 
\beq
{\lambda} = x - {\nu} \frac{{\theta}^{\prime}(u|{\tau})}{{\theta}(u|{\tau})}
\label{eq:aff}
\eeq
makes \eqref{eq:speccurve}
into
\beq
{\CalR}(x, u ) = {\rm Det} \left( x \cdot {\rm Id}_{N} - {\bf {\tilde L}}(u) \right) = 0 
\label{eq:shiftc}
\eeq
with ${\tilde L}(u)$ have a first order pole at $u \in {\BZ}+{\tau}{\BZ}$ with rank one residue.
Therefore, the function ${\CalR}(x,u)$ has, at fixed $x$ only a first order pole at $u = 0$. At fixed $u$ it is still a degree $N$ polynomial in $x$. However, it is not a double-periodic function of $u$ at fixed $x$, rather:
\beq
{\CalR}(x - 2{\pi}{\ii}{\nu}, u + {\tau} ) = {\CalR}(x , u + 1 )  = {\CalR}(x , u ) \, , 
\label{eq:shiftper}
\eeq
Thus ${\CalR} = 0$ is an equation for a curve ${\CalC}$ in the total space $S_{\nu}$ of affine bundle over $T^{*}E_{\tau}$, again with the fiber over $u = 0$ removed.
Note that $S_{\nu}$ has a rich set of holomorphic functions, e.g. any Laurent polynomial in 
\beq
U_{1} = {\exp} \, \left( \frac{x}{\nu} \right) \, , \ U_{2} = {\exp} \, \left( 2\pi \ii u + {\tau} \frac{x}{\nu} \right) 
\label{eq:u1u2f}
\eeq
Thus, for $\nu \neq 0$, outside the fiber over $u=0$ the surface $S_{\nu}$
is isomorphic to the algebraic torus ${\BC}^{\times} \times {\BC}^{\times}$. The curve ${\CalC}$, however, is not algebraic, only analytic, in the $\left( U_1, U_2 \right)$ coordinate system. 

\section{Quantum Krichever curve}

In a very naive way the equation ${\CalR}(x,u) = 0$ can be quantized, for example, by making $x,u$ the generators ${\hat x}, {\hat u}$ of a non-commutative algebra, obeying
\beq
[  {\hat u} , {\hat x}] = {\hbar} \cdot 1
\eeq
where ${\hbar} \in {\BC}$ is a parameter.  It is not obvious how to order ${\hat x}$ and ${\hat u}$
in the complicated function such as ${\CalR}(x, u)$, or, say, $R(x,u)$. However, ${\CalR}(x,u)$ is much better suited for quantization than $R(x, u)$ for the following reason. As we pointed out earlier, the operator ${\bf\tilde L}(u)$ has  a first order pole at $u = 0$ (modulo the lattice ${\BZ} + {\BZ}{\tau}$) with the rank one residue. Thus, 
\beq
{\CalR}(x,u) = \sum_{i=0}^{N} x^{N-i} f_{i}(u)
\eeq
with the coefficients $f_{i}(u)$ having only a first order pole at $u \in {\BZ} + {\BZ}{\tau}$. Of course, the shifted periodicity \eqref{eq:shiftper} means that $f_{i}(u)$'s are not double-periodic meromorphic functions of $u$ (with the exception of $f_{0}(u) \equiv 1$). Now, let us multiply ${\CalR}(x,u)$ by ${\theta}(u|{\tau})$, and define the Fourier coefficients ${\tilde\fR}_{l}(x)$ by:
\beq
{\fR}(x,u) = {\theta}(u|{\tau}) {\CalR}(x,u) = \sum_{l\in {\BZ}} (-1)^{l} e^{2\pi\ii l u} {\qe}^{\frac{l(l-1)}{2}} {\tilde\fR}_{l}(x)
\label{eq:tpol}
\eeq
It is clear from the definition and   \eqref{eq:shiftper} that ${\fR}(x,u)$ is an entire function of $x,u$, degree $N$ polynomial in $x$ at fixed $u$, obeying
\beq
{\fR}(x, u+1) = - e^{2\pi\ii u}\, {\fR}(x-2\pi \ii \nu , u +\tau)  = {\fR}(x , u ) \ .
\eeq
Thus ${\tilde\fR}_{l}(x)$ are all degree $N$ (monic) polynomials in $x$, obeying
\beq
{\tilde\fR}_{l} (x - 2{\pi}{\ii}{\nu}) = {\tilde\fR}_{l+1}(x)
\eeq
In other words 
\beq
{\tilde\fR}_{l}(x) = {\cY} ( {\sf w} - {\nu}l )
\eeq
where $x = 2\pi \ii {\sf w}$, and ${\cY} = {\tilde\fR}_{0}$ is a degree $N$ monic polynomial. 
Now we quantize \eqref{eq:tpol} by replacing $u$ in the exponent by 
\beq
{\hat u} = {\hbar} \frac{\partial}{\partial x} = \frac{\hbar}{2\pi \ii}  \frac{\partial}{\partial \sf w}
\eeq 
We can place $e^{2\pi \ii l {\hat u}}$ to the right of ${\tilde\fR}_{l}(x)$, or we can place it to the left, amounting to a renormalization $\nu \to \nu \pm \hbar$. We can also imagine a dual quantization, where $e^{2\pi\ii l {\hat u}}$ is to the right of ${\tilde\fR}_{l}(x)$ but acts on the left. We shall see both versions realized below. 

The quantum version of the spectral curve can, therefore, be represented as an infinite order difference operator
${\hat\fR}$  with polynomial coefficients, acting on functions of one variable $\sf w$, or its dual. 
The corresponding solutions ${\bf\Psi}$, ${\bf \Psi}^{\vee}$ are the functions of $\sf w$, obeying:
\begin{multline}
 {\hat\fR} {\bf\Psi} ({\sf w})\,  = \, 
\sum_{l \in {\BZ}} (-1)^{l} {\qe}^{\frac{l(l-1)}{2}} {\cY}({\sf w} - {\nu}l) {\bf\Psi} ( {\sf w} + {\hbar} l ) = 0 \\
  {\hat\fR}^{*}  {\bf\Psi}^{\vee} ({\sf w}) \, = \,  \sum_{l \in {\BZ}} (-1)^{l} {\qe}^{\frac{l(l-1)}{2}} {\cY}({\sf w} - {\nu}l-{\hbar}l) {\bf\Psi}^{\vee} ( {\sf w} - {\hbar} l ) = 0\\
\label{eq:qceq}
\end{multline}

The quantum version of the spectral curve can, equivalently, be represented in $u$-variable, with
\beq
{\widehat{\sf w}} = \frac{\hbar}{2\pi\ii} \frac{\partial}{\partial u}
\eeq
as the $N$'th order meromorphic differential operator on $E_{\tau}$, with a regular singularity at $u=0$,  a $PGL_{N}$-oper. The corresponding Fourier transforms
\beq
\begin{aligned}
& {\chi}_{\alpha}(u) = \sum_{{\sf w} \in {\alpha} + {\hbar}{\BZ}} e^{\frac{2\pi \ii {\sf w}}{\hbar} u} \, {\bf\Psi} ( {\sf w} ) \\
& {\chi}_{\alpha}^{\vee} (u) = \sum_{{\sf w} \in {\alpha} + {\hbar}{\BZ}} e^{-\frac{2\pi \ii {\sf w}}{\hbar} u} \, {\bf\Psi}^{\vee} ( {\sf w} ) \end{aligned}
\label{eq:chifrompsi}
\eeq
where $\alpha$ is chosen so that the series on the right hand side of \eqref{eq:chifrompsi} converge are the local solutions of the oper. They should play a r{\^o}le in the quantum separation of variables for eCM and its spectral dual $Y(\widehat{\mathfrak{gl}}(1))$-based spin chain, analogously to the genus zero ($\mathfrak{sl}_{r+1}$ spin chain) case studied in \cite{Jeong:2023qdr}. Also, the construction of \cite{EFK} seems to generalize to genus one case, giving rise to the $L^2$,  or $z, {\bar z}$-version of the elliptic Calogero-Moser system, where the wave-function of the $N$-body system is an $L^2$-function on $E_{\tau}^{N}/S(N)$, which is a common eigenvector of the Hamiltonians of the conventional, meromorphic, eCM and its complex conjugate. The parameters $\alpha$ and the coefficients of the polynomial ${\cY}$ would have to be fixed by a requirement of the $PGL_{N}({\BR})$-holonomy of the oper. The ${\CalN}=2^{*}$ counterpart of this construction seems to be placing the theory on extremely elongated four dimensional ellipsoid with a surface defect filling up a two dimensional equatorial section thereof. 

We shall return to this picture elsewhere.

\section{$T-Q$-equation}

Functional Bethe ansatz \cite{Sklyanin:1991ss} (see \cite{Dr1, FT, TYBE, Sklyanin} for the reviews of the foundational work) approach to $\mathfrak{sl}_{2}$-spin chains or quantum periodic $N$-body Toda chain uses the following $T-Q$-equation \cite{Baxter,Reshet}:
\beq
P_{+}({\sf w}-2{\hbar}) {\bQ}({\sf w}+\hbar) + {\qe} P_{-}({\sf w}+2{\hbar}) {\bQ}({\sf w}-\hbar) = (1+{\qe}) {\bT}({\sf w}) {\bQ}({\sf w})
\label{eq:tqsl2}
\eeq
with degree $N$ monic polynomial ${\bT}({\sf w})$ and entire function ${\bQ}({\sf w})$. 
The Eq. \eqref{eq:tqsl2} describes the eigenvalues ${\bf T}({\sf w})$ of the transfer matrix ${\hat{T}}({\sf w})$.  There is a {\em spectral dual} Lax operator ${\bf L}(z)$ of a particular case of $\mathfrak{sl}_{N}$ Gaudin-Garnier model, whose eigenvalues belong to the ${\sf w}$ line. 
The equation \eqref{eq:tqsl2} can be derived from the Yangian version of the $q$-characters of $\mathfrak{sl}_{2}$ \cite{FR}, from the ${\ve}_{2} \to 0$ limit of the $qq$-characters of the $A_1$-type ${\CalN}=2$ superconformal gauge theory (= $N_{f}=2N_{c}$ super-QCD) \cite{N3}, it can also be viewed as a quantum version of the Seiberg-Witten curve of the $A_1$ type gauge theory (for the topological recursion approach to quantum curves of that type see \cite{Chekhov:2012rd}, and for the extension to the local CY geometries \cite{Coman:2022igv}). 
In this paper we define (and solve) an ${\widehat{\mathfrak{gl}(1)}}$-analogue of \eqref{eq:tqsl2}.

\section{Partitions, characters, and contents}

  In what follows $\Lambda$ denotes the set of all partitions, i.e. non-increasing sequences 
  \beq
  {\Lambda} = \Biggl\{ \, {\lambda} \, | {\lambda} = \left( {\lambda}_{1} \geq {\lam}_{2} \ldots \geq {\lam}_{{\ell}({\lambda})} > 0 = {\lambda}_{{\ell}({\lambda})+1} \right) \Biggr\} 
  \eeq 
 The size and length maps
 \beq
 | \cdot |\, , {\ell} \, : \, {\Lambda} \longrightarrow {\BZ}_{\geq 0} \, , \qquad |{\lam}| \equiv {\lam}_{1} +\ldots + {\lam}_{\ell ({\lam})} \ , 
 \eeq
 and the ``transposition'' map: ${\lam} \mapsto {\lam}^{t}$, ${\ell}({\lam}^{t}) = {\lam}_{1}$, $|{\lam}^{t}| = |{\lam}|$, best described by 
 viewing partitions as finite subsets $\lambda \subset {\BZ}_{>0} \times {\BZ}_{>0}$:
 \beq
 {\lambda} = \Biggl\{  {\square} = (i,j) \,  | \, i, j \geq 1\, \ 1 \leq j \leq {\lam}_{i}\, , \ 1 \leq i \leq {\lam}_{j}^{t} \, \Biggr\} 
 \label{eq:psubset}
 \eeq
 Define the {\em contents} of $\square$:
  \beq
  {\xi}_{\square}= {\hbar}\left( i-j + n (j-1) \right) \, , \ {\upsilon}_{\square} = {\hbar} \left( n(j-i) + 1-j \right)  \, , \  {\zeta}_{\square} = {\hbar} \left( i-1 - n (j-1) \right)
  \label{eq:contents}
  \eeq
  Define the {\em boundary sets} ${\Gamma}^{+}_{\lambda}$, ${\Gamma}^{-}_{\lambda}$ by: 
  \beq
  \sum_{(i,j) \in {\Gamma}^{+}_{\lambda}} \langle i-1, j-1\rangle -  \sum_{(i,j) \in {\Gamma}^{-}_{\lambda}} \langle i, j\rangle = \sum_{i=1}^{{\ell}({\lambda})+1} \langle i-1, {\lambda}_{i} \rangle  -   \sum_{i=1}^{{\ell}({\lambda})} \langle i, {\lambda}_{i}\rangle 
  \eeq
  where $\langle i , j \rangle$ is any linear function $\langle i, j \rangle = (i-1) + f(j-1)$, for $f \in {\mathbb C}\backslash{\mathbb Q}$. 
  
 Finally, define the character
 \beq
 {\chi}_{ab}({\lambda}) = \sum_{{\square} \in {\lam}}  q_{a}^{i-1} q_{b}^{j-1}
 \label{eq:char}
 \eeq
 for $a,b \in \{ 1 , 2, 3, 4 \}$

\section{$Q$-functions, $\sf X,Y,Z$-observables, and ${\cX}, {\cY}, {\cZ}$-characters}  

Supersymmetric gauge theories produce \cite{N3} (see Appendix $\bf A$ for some details) entire functions $Q({\sf w})$, which, in the limit ${\ve}_{2} \to 0, {\ve}_{1} = {\hbar}$ of \cite{N2} solve {\em Bethe equations}
\beq
\frac{Q({\sf w}+{\hbar})Q({\sf w}-{\hbar}n)Q({\sf w}+{\hbar}(n-1))}{Q({\sf w}-{\hbar})Q({\sf w}+{\hbar}n)Q({\sf w}+{\hbar}(1-n))} = - {\qe}\, , \ \forall {\sf w} \in Q^{-1}(0) \ . 
\label{eq:bae}
\eeq 
Define 
\beq
{\mathsf{X}}({\sf w}) = \frac{Q({\sf w})}{Q({\sf w}+{\hbar}n)} \, , \ {\sf Y}({\sf w}) = \frac{Q({\sf w})}{Q({\sf w}-{\hbar})}\, , \ {\sf Z}({\sf w}) = \frac{Q({\sf w})}{Q({\sf w}+{\hbar}(1-n))} \, . \eeq
Then $U(N)$ ${\cN}=2^{*}$ theory with adjoint hypermultiplet of mass ${\hbar}n$ on $\Omega$-deformed ${\bR}^{2}_{\hbar} \times {\BR}^{1,1}$ is characterized by the property that 
\beq
{\sf Y}({\sf w}) \sim {\sf w}^{N} +\ldots \,  {\rm at} \ {\sf w} \to \infty
\label{eq:largewasy}
\eeq 
(it is called the $Y$-observable in \cite{N3}). Define the following triplet of observables built out of $\mathsf{X}, \mathsf{Y}, \mathsf{Z}$, respectively: 
\beq
\begin{aligned}
&   {\cX}({\sf w}) \,  = \, \sum_{\lambda} {\qe}^{|{\lam}|}  \, \frac{\prod_{{\square} \in {\Gamma}^{+}_{\lambda}} {\sf X} ( {\sf w} - {\hbar}n + {\xi}_{\square})}{\prod_{{\square} \in {\Gamma}^{-}_{\lambda}} {\sf X} ( {\sf w} + {\xi}_{\square})}\, ,  \\
& {\cY}({\sf w}) \,  = \, \sum_{\lambda} {\qe}^{|{\lam}|} \, \frac{\prod_{{\square} \in {\Gamma}^{+}_{\lambda}} {\sf Y} ( {\sf w} + {\hbar} + {\upsilon}_{\square})}{\prod_{{\square} \in {\Gamma}^{-}_{\lambda}} {\sf Y} ( {\sf w} + {\upsilon}_{\square})} \, ,  \\
&  {\cZ}({\sf w})  \,  = \, \sum_{\lambda} {\qe}^{|{\lam}|} \, \frac{\prod_{{\square} \in {\Gamma}^{+}_{\lambda}} {\sf Z} ( {\sf w} + {\hbar}(n -1) + {\zeta}_{\square})}{\prod_{{\square} \in {\Gamma}^{-}_{\lambda}} {\sf Z} ({\sf w} + {\zeta}_{\square})}   \\
\end{aligned}
\label{eq:xyz}
\eeq
Then ${\cY}({\sf w})$ is a degree $N$ polynomial in ${\sf w}$, as \eqref{eq:bae} guarantees the absence of poles, and the asymptotics \eqref{eq:largewasy} fixes the rest (it is a particular, ${\hat A}_{0}$,  case of the main theorem of \cite{N3}). In fact, ${\cY}({\sf w})$ is the ${\ve}_{2} \to 0$ limit of the the fundamental $qq$-character of ${\hat A}_{0}$-theory,  introduced in \cite{N3}. It can be related to the $q$-character of \cite{FR} for the $\widehat{\mathfrak{gl}}(1)$-algebra. Our main \emph{new} result is the set of three
\medskip
\section{Bilinear equations}

\beq
\begin{aligned} 
{\cY}\star {\cX} ({\sf w}) \equiv \sum_{l \in {\BZ}} (-1)^{l} {\qe}^{\frac{l(l+1)}{2}} \, {\cY} ({\sf w} -  {\hbar}n  (l+1)) {\cX}({\sf w}- {\hbar} l)  = 0   \\
{\cY}\star {\cZ} ({\sf w}) \equiv \sum_{l \in {\BZ}} (-1)^{l} {\qe}^{\frac{l(l+1)}{2}} \, {\cY} ( {\sf w} + {\hbar}(n-1)(l+1) ) {\cZ} ({\sf w} - {\hbar} l )  = 0 \\
{\cX}\star {\cZ} ({\sf w}) \equiv \sum_{l \in {\BZ}} (-1)^{l} {\qe}^{\frac{l(l+1)}{2}} \, {\cX} ({\sf w} + {\hbar}(n-1)(l+1)) {\cZ} ( {\sf w} + {\hbar} n l )  = 0
\end{aligned}
\label{eq:maineqs}
\eeq
A quick comparison with \eqref{eq:qceq} reveals ${\cX}$ and ${\cZ}$ 
are the dual solutions of the quantum spectral curve  \eqref{eq:shiftc}. Having these solutions is a step towards the quantum separation of variables \cite{Sklyanin:1995bm} for the elliptic CM system. 

\section{Discussion: gauge theory and integrable system}

Gauge theory approach to elliptic Calogero-Moser system was started in \cite{GN3}. It was shown there that the phase space and the Hamiltonians of the many-body system can be obtained by Hamiltonian reduction from an infinite-dimensional phase space, which is the space of two-dimensional gauge fields and adjoint-valued one-forms (mathematically called Higgs fields, although they were introduced and studied by N.~Hitchin in \cite{Hitchin}, the construction of \cite{GN3} being a degenerate version \cite{NHo}). In this realization the classical and quantum evolution of the many-body system is identified with that of some $2+1$ dimensional gauge theory. The proposal \cite{Donagi:1995cf} identified the family of spectral curves found in \cite{GN3} (which thanks to \cite{Krichever80} are the spectral curves of elliptic Calogero-Moser system) with the family of Seiberg-Witten curves of ${\CalN}=2^{*}$ theory with gauge group $SU(N)$ (an earlier paper \cite{Gorsky:1995zq} pointed out a connection between the Seiberg-Witten proposal for the geometry of vacua of ${\CalN}=2$ theories and integrable systems).  The analysis \cite{DHoker:1996pva} allows to extract the low-instanton number non-perturbative corrections to the low-energy effective action from the careful expansion of the periods of a degenerating family of curves. The bottom up approach \cite{N2}, starting with supersymmetric gauge theory, computing the effects of the non-perturbative 
dynamics on the geometry of its moduli space of vacua, produced a two-parameter generalization of the Seiberg-Witten geometry. It was pointed out that these generalizations might lead to the quantization of the integrable system, as well as other deformations. In \cite{NO, NP1} important steps were made in showing that indeed, for ${\CalN}=2^{*}$ theory (among others) the gauge theory computation produces the spectral curve of the elliptic Calogero-Moser system. 

Our present achievement is to find, within gauge theory, the observables whose expectation values play the role of the {\it single-particle wave-functions} for the quantum version of the spectral curve. We build them using the technique of {\it folded instantons}, introduced in \cite{N3}. Again, we stress that our results are derived, not conjectured. In this way we made independent checks of many string duality based claims 
e.g. \cite{Aganagic:2003qj, AGT, G, KKV, BKK}.  The proper embedding of folded instanton observables in gauge theory remains an interesting open problem (see \cite{Nekrasov:2016gud}
for string theory considerations). 

Likewise, the meaning of the sums over $l$ in \eqref{eq:maineqs} begs for a gauge theory explanation. Our conjecture is that just like the ${\cY}$-observable can be interpreted as a partition function of an auxiliary four dimensional theory living on a copy of ${\bR}^{4}$ transverse to the physical spacetime ${\bR}^{4}$ in some ambient eight-dimensional special holonomy geometry, the sum in ${\fR}(x, u)$ is the partition function of a similar $U(1)$ theory placed in the Taub-Nut background. The sum over $l \in {\BZ}$ is the sum over the fluxes of the abelian gauge field, which is supported by the $L^2$-normalizable self-dual harmonic two-form. One needs to find a Taub-Nut $\times {\BR}^{4}$ analogue of the moduli space of crossed instantons. The ADHM construction for instantons on Taub-Nut spaces has been beautifully generalized in \cite{Cherkis:2009hpw}. 

Of course, the relation between \eqref{eq:qceq} and the quantum spectral curve of elliptic Calogero-Moser system is not very direct. We simply match the families of curves, and their non-commutative versions. However we can claim the match of the coordinate systems, the coordinates $z_{1}, \ldots, z_{N}$ of the particles having gauge-theoretic significance. We know from \cite{N3} the regular surface defect of ${\CalN}=2^{*}$ theory obeys eCM Schr{\"o}dinger equation, with $z_{i}-z_{i+1}$ being the complex K{\"a}hler moduli of the surface defect. As we said in the beginning, we are discussing a simpler version of eCM, defined in complex-analytic terms, so there are separate ``in'' and ``out'' wave-functions, related by ${\hbar} \mapsto -{\hbar}$ transformations. A short form of   \eqref{eq:maineqs} reads
\begin{multline}
0 =  \left( \sum_{l \in {\BZ}} (-1)^{l} {\qe}^{\frac{l(l+1)}{2}} \, {\cY} ({\sf w} -  {\hbar}n  (l+1)) e^{-{\hbar}l {\partial}_{\sf w}} \right) {\Psi} \, , \\ 
0 =  \left( \sum_{l \in {\BZ}} (-1)^{l} {\qe}^{\frac{l(l+1)}{2}} \, e^{{\hbar}l {\partial}_{\sf w}}  {\cY} ( {\sf w} - {\hbar}n (l+1) ) \right) {\Psi}^{\vee}\, , 
\end{multline}
where ${\Psi}({\sf w}) = {\cX}  ({\sf w}) \,  , \ {\Psi}^{\vee} ({\sf w}) = {\cZ} ({\sf w} + {\hbar} (1-n) ) $. The quantum separation of variables which would express the regular surface defect as an integral transform of a product of the single-particle wave-functions, which should be the ${\Psi} \sim {\cX}$ or ${\Psi}^{\vee} \sim {\cZ}$-observables. 

The solutions ${\cX}$ and ${\cZ}$ are the ${\hat A}_{0}$ analogues of the two independent (over the quasiconstants, i.e. $\hbar$-periodic functions) solutions of the $T-Q$ equations. The third bilinear equation ${\cX} \star {\cZ} = 0$ is the analogue of the very important
\emph{quantum Wronskian relation} (see the Appendix $\bf B$). 

In the forthcoming publication \cite{GKN} we shall establish another detail in the gauge theory-integrable system package dictionary: the Lax operators as gauge theory observables. Our main equations
\eqref{eq:maineqs} will get a matrix-vector dress explaining the origin of so many beautiful \emph{zero-curvature representations} in the theory of integrable systems. The manipulation 
\eqref{eq:tpol} leading from $\CalR$ to $\fR$ will also get a matrix version, generalizing that of \cite{GZ, VZ}. The story of the ${\CalN}=2^{*}$ (${\hat A}_{0}$) theory we discussed generalizes to the ${\hat A}_{r}$-type quiver theories, with \eqref{eq:maineqs} mapping, upon a Fourier transform, to $PGL_N$-opers on elliptic curve $E_{\tau}$ with $r+1$ regular singularities with minimal monodromies around the poles (another limit, ${\ve}_{1} + {\ve}_{2} \to 0$ of this theory was studied recently in \cite{Bonelli:2019yjd}). 

The quantum separation of variables for eCM is still around the corner...

Our equations should also play a r{\^o}le in the solution of the quantum intermediate long-wave equation, a $N \to \infty$ counterpart of eCM system, related to the six dimensional gauge theories and sigma models on instanton moduli spaces \cite{OP2004}, as well as to the integrable structure of two dimensional conformal field theory \cite{Bazhanov:1994ft}, and quantum hydrodynamics \cite{AO:2013,Litvinov:2013zda}.

\section{Acknowledgments}

We enormously benefited from discussions and collaboration with I.~Krichever in 1995-2022. 
We also thank S.~Cherkis and S.~Grushevsky for discussions. 
Research was partly supported by NSF PHY Award 2310279. Any opinions expressed are solely our own and do not represent the views of the National Science Foundation.

\section{Appendix $\bf A$. Derivation of the main bilinear equations}

In this section we use the notations of \cite{N3}. Let ${\ve}_{1}, {\ve}_{2}, {\ve}_{3}, {\ve}_{4} \in {\BC}$
sum to zero, 
\beq
{\ve}_{1}+ {\ve}_{2}+ {\ve}_{3}+ {\ve}_{4} = 0
\eeq
and let $q_{a} = e^{{\ve}_{a}}$, $P_{a} = 1-q_{a}, P_{ab} = P_{a}P_{b}$, $a, b, = 1, \ldots, 4$. The ${\cY} \star {\cX}$ of \eqref{eq:maineqs} is the ${\ve}_{2} \to 0$ limit of the sum over $l$ of
\beq
(-1)^{l} {\qe}^{\frac{l(l+1)}{2}} \times Z_{{\ve}_{1}, {\ve}_{2}=0, {\ve}_{3}, {\ve}_{4}}( N_{ab} )\, , 
\eeq 
the partition function of the gauge origami theory with
\beq
{\ve}_{1} = {\hbar}\, , \ {\ve}_{3} = - {\hbar}n \, , \ {\ve}_{4} = {\hbar}(n-1)
\label{eq:e1e3e4}
\eeq 
and the Chan-Paton characters
\beq
N_{12} = \sum_{\alpha = 1}^{N} e^{a_{\alpha}}\, , \ N_{34} = e^{{\sf w} + {\ve}_{3} (l+1)}\, , \ 
N_{14} = e^{{\sf w} - {\ve}_{1} l} \, , 
\eeq
the ${\cY} \star {\cZ}$ corresponds to the origami with Chan-Paton characters
\beq
N_{12} = \sum_{\alpha = 1}^{N} e^{a_{\alpha}}\, , \ N_{34} = e^{{\sf w} + {\ve}_{4} (l+1)}\, , \ 
N_{13} = e^{{\sf w} - {\ve}_{1} l} \, , 
\eeq
finally ${\cX} \star {\cZ}$ corresponds to the origami with
\beq
N_{12} = \sum_{\alpha = 1}^{N} e^{a_{\alpha}}\, , \ N_{14} = e^{{\sf w} + {\ve}_{4} (l+1)}\, , \ 
N_{13} = e^{{\sf w} - {\ve}_{3} l} \, ,
\eeq
We prove the ${\cY} \star {\cX} = 0$ equation from \eqref{eq:maineqs} by constructing a involution 
\beq
{\Upsilon}: {\Lambda} \times {\Lambda} \times {\BZ}  \longrightarrow  {\Lambda} \times {\Lambda} \times {\BZ} \ , 
\eeq
sending
$({\lambda}, {\mu},  l)$ to $({\tilde\lambda}, {\tilde\mu}, {\tilde l})$, such that:
\beq
\begin{aligned}
1) & \qquad |l - {\tilde l}| = 1 \\
2) & \qquad P_{1} S_{34} (l, {\lambda} )  + P_{3} S_{14}  (l, {\mu} ) = 
P_{1} S_{34} ({\tilde l}, {\tilde\lambda} )  + P_{3} S_{14}  ({\tilde l}, {\tilde\mu} ) \\
\end{aligned}
\label{eq:cancel}
\eeq
the latter equation implying
\beq
\frac{l(l+1)}{2} + | {\lambda}  | + | {\mu}  |  = \frac{{\tilde l}({\tilde l} + 1)}{2} + | {\tilde\lambda} | + | {\tilde\mu}|  \label{eq:sizes}
\eeq
where 
\begin{multline}
S_{34}(l, {\lambda}) = e^{{\sf w}} q_{3}^{l+1} \left( 1 - P_{34} {\chi}_{34} ({\lambda}) \right)\, , \\ 
S_{14}  (l, {\mu} ) = e^{{\sf w}} q_{1}^{-l} \left( 1 - P_{14} {\chi}_{14} ({\mu}) \right)\, , \end{multline}
The involution acts by:
 \beq
 {\Upsilon} ({\lam}, {\mu}, l ) = \begin{cases} \left( {\tilde\lam}, {\tilde\mu}, l-1 \right) \, & , \qquad {\tilde\lam}_{1} = {\mu}_{1} + l\, , \ {\tilde\lam}_{i+1} = {\lam}_{i} \,  , \ {\mu}_{i+1} = {\tilde\mu}_{i}, \ i \geq 1\ , 
  \\
  & \qquad\qquad {\rm when}\ \ {\mu}_{1} + l \geq {\lam}_{1} \ , \\
 \left( {\hat\lam}, {\hat\mu}, l+1 \right) \, & , \qquad  {\hat\mu}_{1} = {\lam}_{1} - l - 1\, , \  {\hat\lam}_{i} = {\lam}_{i+1}\, , \ {\hat\mu}_{i+1} = {\mu}_{i} \, , \ i \geq 1 \ , 
 \\
 & \qquad\qquad {\rm when}\ \ {\mu}_{1} + l < {\lam}_{1} \ .
 \end{cases}
 \label{eq:involution}
 \eeq
Let us check \eqref{eq:cancel} in the case ${\mu}_{1} + l \geq {\lam}_{1}$, i.e. ${\tilde l} = l-1$. In that case, the second equation in \eqref{eq:cancel} reads
\beq
 \frac{1  - q_{4}^{l}}{1-q_{4}}  +     q_{3} {\chi}_{34} ( {\lambda} ) + q_{4}^{l} {\chi}_{14} ({\mu}) =   {\chi}_{34} ({\tilde\lambda} )  +  q_{4}^{l}  q_{1} {\chi}_{14} ( {\tilde\mu}) 
 \label{eq:matchacha}
 \eeq
 Let us assume $l \geq 1$ (the case $l \leq 0$ is analogous) and expand the left hand side of \eqref{eq:matchacha} in $q_{1}, q_{4}$:
 \beq
 \sum_{i=1}^{{\ell}({\lambda})}q_{1}^{-i} \sum_{j=1}^{{\lam}_{i}} q_{4}^{j-i-1} + \sum_{j=1}^{{\mu}_{1}+l} q_{4}^{j-1} + 
 \sum_{i=2}^{{\ell}({\mu})} q_{1}^{i-1} \sum_{j=l+1}^{{\mu}_{i} + l} q_{4}^{j-1}
 \label{eq:lhsm}
 \eeq
 while the right hand side of \eqref{eq:matchacha} expands, similarly, as
 \beq
 \sum_{i=1}^{{\ell}({\tilde\mu})} q_{1}^{i} \sum_{j=l+1}^{{\tilde\mu}_{i}+l} q_{4}^{j-1} + 
 \sum_{i=1}^{{\ell}({\tilde\lam})} q_{1}^{1-i} \sum_{j=1}^{{\tilde\lam}_{i}} q_{4}^{j-i}
 \eeq
 Comparing the $q_{1}^{0}$, $q_{1}^{<0}, q_{1}^{\geq 0}$ terms, respectively, we get the precise match, given \eqref{eq:involution}.

 \section{Appendix $B$. Limit to periodic Toda and relation to earlier work}

 The Toda limit, i.e.
 \beq 
 n \to \infty\, , \qe \to 0\, , 
 \label{eq:nqe}
 \eeq
 while keeping finite 
 \beq
 {\Lambda}^{2N} = {\qe} (-1)^{N} \left( {\hbar}n \right)^{2N}
 \eeq
 is a bit delicate, as
 far as the ${\cX},{\cZ}, {\sf X}, {\sf Z}$ functions are concerned. The $Q$-function has a smooth limit, with the Bethe equations \eqref{eq:bae} going over to 
\beq
\frac{Q({\sf w}+{\hbar})}{Q({\sf w}-{\hbar})} = - {\Lambda}^{2N}\, , \ \forall {\sf w} \in Q^{-1}(0) \ . 
\label{eq:bae-toda}
\eeq 
More precisely, this is the limit of $Q({\sf w})$ at fixed $\sf w$. On the other hand, $Q$ with the shifted arguments have somewhat singular limits, which we need to account for:
\beq
Q({\sf w} - {\hbar} n l) \sim\  (n l )^{\frac{N}{2}} \, e^{N n l}\, \left( - {\hbar} n  l \right)^{\frac{N {\sf w}}{\hbar}- N n l} \times \left(1 + O(n^{-1}) \right)
\label{eq:qlw} 
\eeq  
Specifically, we shall need \eqref{eq:qlw} for $l=-1,1$ and a limit of the product
\beq
{\qe}^{\frac{\sf w}{\hbar} + p} Q({\sf w} + {\hbar} (p- n)) Q ({\sf w} + {\hbar}(n-1+p)) \sim \frac{(-1)^{N \left( n + \frac{1}{2} \right)}}{\hbar^{N}} {\Lambda}^{2N \left( \frac{\sf w}{\hbar}+  p \right)} 
\label{eq:prodlim}
\eeq
The series for ${\cX}({\sf w}), {\cZ}({\sf w})$, in our limit, receive contributions only from partitions ${\lambda} = (1^{p})$, for $p = 0, 1, \ldots$, giving:
\beq
\begin{aligned}
& {\cX} ({\sf w}) \sim  {\tilde Q}({\sf w}-{\hbar}) \, , \\
& {\cZ} ({\sf w}) \sim  {\tilde Q}({\sf w}-{\hbar})\, , \\
\end{aligned}
\label{eq:cxztoda}
 \eeq
 where $\sim$ means up to a factor $e^{c_{1}(n){\sf w} + c_{0}(n)}$, 
 \beq
 {\tilde Q}({\sf w}) = {\Lambda}^{2 N \frac{\sf w}{\hbar}} \, Q ( {\sf w} ) \sum_{p=0}^{\infty} \frac{{\Lambda}^{2N p}}{Q ( {\sf w} + {\hbar}p) Q ( {\sf w} + {\hbar} (p+1) )} \ .
 \label{eq:dualq}
 \eeq
 The ${\hbar}n$-shifted ${\cX}({\sf w}), {\cZ}({\sf w})$ functions have $Q$-function as its principal asymptotics:
 \beq
 {\cX}({\sf w} + {\hbar}n) \sim Q({\sf w}) \, , \ {\cZ}({\sf w} + {\hbar}(1-n)) \sim Q({\sf w})
 \label{eq:qfromxz}
 \eeq
 Finally, the series for ${\cY}({\sf w})$ retains only two terms in the limit \eqref{eq:nqe}:
 \beq
 {\cY}({\sf w}) = {\mathsf{Y}}({\sf w} + {\hbar} ) + \frac{{\Lambda}^{2N}}{{\mathsf{Y}}({\sf w})}
 \label{eq:yy}
 \eeq
 Either by recalling the relation between $Q({\sf w})$ and $\mathsf{Y}({\sf w})$, or by taking the limit of \eqref{eq:maineqs} we obtain 
 the \emph{T-Q equation} (cf. \cite{Jeong:2023qdr,Poghosyan:2016mkh, N3}) , obeyed both by $Q({\sf w})$
 and ${\tilde Q}({\sf w})$:
 \beq
 \begin{aligned}
& Q({\sf w}+{\hbar}) + {\Lambda}^{2N} Q ({\sf w} - {\hbar}) = {\cY}({\sf w})  Q({\sf w}) \\
&  {\tilde Q}({\sf w}+{\hbar}) + {\Lambda}^{2N} {\tilde Q} ({\sf w} - {\hbar}) = {\cY}({\sf w}) {\tilde Q}({\sf w}) \\
\end{aligned} \label{eq:dualtq}
 \eeq
in complete agreement with \eqref{eq:bae-toda}. 
 Note that \eqref{eq:dualtq} can also be obtained by a limit of \eqref{eq:tqsl2}. 
 
 The Toda limit \eqref{eq:nqe} of \eqref{eq:maineqs} gives \eqref{eq:dualtq} and the Wronskian relation (cf. \cite{Kozlowski:2010tv})
 \beq
 {\tilde Q}({\sf w})Q({\sf w}+{\hbar}) -  Q({\sf w}) {\tilde Q}({\sf w}+{\hbar}) = {\Lambda}^{\frac{2N{\sf w}}{\hbar}}
 \label{eq:wronsk}
 \eeq
Both $Q$ and ${\tilde Q}$ functions prominently feature (in disguise) in the separation of variables for the quantum Toda system (the Ref. \cite{Kharchev:2000yj} used determinant representation of Gaudin-Pasquier \cite{GP}, which can be related to our $Q$, ${\tilde Q}$'s). 

 We leave the verification
 to the reader as an exercise in application of \eqref{eq:cxztoda}, \eqref{eq:prodlim}, \eqref{eq:qlw}.

\end{document}